\documentstyle[prd,aps,floats]{revtex}
\begin{document}
\draft

\input epsf \renewcommand{\topfraction}{0.8} 
\twocolumn[\hsize\textwidth\columnwidth\hsize\csname 
@twocolumnfalse\endcsname

\title{Interesting consequences of brane cosmology} 
\author{Anupam Mazumdar}
\address{Astrophysics Group, Blackett Laboratory, Imperial College London,
 SW7 2BZ,~U.~K.}
\date{\today} 
\maketitle
\begin{abstract}
We discuss cosmology in four dimensions within a context of brane-world scenario.
Such models can predict chaotic inflation with very low reheat temperature depending
on the brane tension. We notice that the gravitino abundance is different in the
brane-world cosmology  and by tuning the brane tension it is possible
to get extremely low abundance. We also study Affleck-Dine baryogenesis
in our toy model. 
\end{abstract}


\vspace*{-0.6truecm}
\vskip2pc]



Recently there has been a renewed interest in perceiving the four 
dimensional world which is in a form of three dimensional hypersurface along with time
embedded in a higher dimensional space-time. Such a claim has a pedigree from
strongly coupled sector of $E_8 \times E_8$ hetrotic string theory which can be 
described by a field theory living in a $11$ dimensional 
space-time \cite{horava}.
The $11$ dimensional world is comprised of two $10$ dimensional hypersurfaces embedded
on an orbifold fixed points, where fields are assumed to be confined to the hypersurfaces 
which are known to be $9$ branes in this scenario. After compactifying the $11$ 
dimensional theory on a Calabi-Yau three fold, one obtains an effective $5$ 
dimensional theory \cite{lukas0},
which has a structure of two $3$ branes situated on the orbifold boundaries.
The theory allows $N=1$ supergravity with gauge and chiral multiplets
on the two $3$ branes. Thus, it is possible to get phenomenologically interesting $N=1$
supergravity in four dimensions from the hetrotic string theory. The low energy 
theory in four dimensions also allows a rich cosmological implications and in recent
past some attempts have been made to understand the cosmology \cite{lukas1}.

In this paper we consider a very simple toy model in $5$ dimensions and we assume 
that we reside in one of the two $3$ branes which are separated by a distance. In this 
set-up it has been realized that the effective $4$ dimensional cosmology is non-conventional
\cite{binetruy}. The Friedmann equation is modified due to localization of
the fields on the brane and also due to presence of the second brane.
The extra  $5$th dimension is assumed to have orbifold symmetry $y =-y$, and static in our case. 
The main goal of this paper is to point out some of the interesting implications
of the brane cosmology taking place at energy scales below four dimensional Planck mass 
and above the nucleosynthesis scale. It is fairly well recognized
that the root of most of the nagging problems of presently observed Universe 
have some relation to the early Universe. We mention here two of them. The present 
Universe seems to be extremely flat, isotropic and homogeneous. A small inhomogeneity is 
measured to be one part in $10^5$ by Cosmic Explorer Background (COBE) satellite, and the 
second startling observation is that the present observable Universe has a small baryon 
asymmetry which is noted to be roughly one part in $10^{10}$, measured from the abundances 
of light elements synthesized at the time of nucleosynthesis.  
A small inhomogeneity of the Universe can be explained by quantum fluctuations
of the scalar fields during inflation. While the observed baryon asymmetry can 
also be explained quite elegantly in the early Universe because of the presence 
of a preferred time and the expansion of the Universe 
which leads to out-of-equilibrium decay of massive particles via explicit CP
violation interactions. In this paper we will consider one such example of baryogenesis in
supersymmetric theories which is known as Affleck-Dine (AD) mechanism \cite{ad}.
Strictly speaking we will be treating the branes as 
hypersurfaces. We will be assuming that initial configuration of the branes are 
supersymmetric and due to some known or unknown reasons supersymmetry is broken at 
a suitable scale to solve the hierarchy problem in the Planck brane where we reside. Regarding this we are
assuming that our set-up has two branes with opposite brane tensions 
and a negative bulk cosmological constant. In this respect our discussion could as well be 
generalizable to the configuration where gravity can be localized on the Planck brane
\cite{lisa} and some attempts have been made to supersymmetrize the two branes
\cite{bagger}. However, some formal aspects of supersymmetrizing infinitely thin branes 
are still under extensive study \cite{kallosh,stelle}, and more recently \cite{duff}.

A simple isotropic and homogeneous cosmology can be described by the expansion
parameter known as the Hubble parameter. It has been noticed that the two branes with
opposite brane tensions can cancel the negative bulk cosmological constant \cite{cgs}
to give rise to a simple modification to the expansion equation. The Friedmann equation
in the Planck brane is given by
\begin{eqnarray}
\label{main0}
H^2=\frac{8\pi}{3M_{\rm p}^2}\rho\left[1+\frac{\rho}{2\lambda}\right]\,,
\end{eqnarray}
where $\rho$ is the energy 
density of the matter stuck to the brane. The brane tension 
$\lambda$ relates the four dimensional Planck mass $M_{\rm p}\approx 10^{19}{\rm GeV}$ 
to the five dimensional Planck scale $M_{5}$ via
\begin{eqnarray}
\label{main1}
M_{\rm p}=\sqrt{\frac{3}{4\pi}}\left(\frac{M_{5}^2}{\sqrt{\lambda}}\right)M_{5}\,.
\end{eqnarray}  
It is noticeable from Eq.~(\ref{main0}) that there is an extra contribution 
to the right-hand side of the Friedmann equation. If we demand that successful 
nucleosynthesis occurs then the second term proportional to $\rho^2$ has to 
play a negligible role at a scale $\sim  {\cal O}(\rm{ MeV})$, corresponding 
to the era of Big Bang nucleosynthesis. Thus we have to assume that the modified 
Friedmann equation paves the usual term in the right-hand side of Eq.~(\ref{main0}), 
which is just linear in energy density. This naturally leads to constraining the 
brane tension as $\lambda > (1 \rm{MeV})^4$ \footnote{A more stringent constraint
on the brane tension has been obtained in Ref.~\cite{lisa1}, by considering the 
fifth dimension to be non-compact and also from the validity of the Newtonian 
gravity in $3+1$ dimensiuons on scales more than $1$mm, which lead to 
constraning the brane tension $\lambda >(100 {\rm GeV})^4$.}. This means that 
the Universe evolves exactly in a familiar fashion even in the presence of branes
at energy scales lower than a MeV. However, there could be a significant 
departure from the 
usual lore at very high energies, especially when $2 \lambda < \rho$. In this
regime the expansion rate of the Universe is certainly dominated by the $\rho^2$ term 
in the right-hand side of Eq.~(\ref{main0}). Our aim is to illustrate that perhaps
we can accommodate the non-conventional term in Eq.~(\ref{main0}) for solving some
of the problems, such as excess gravitino production during reheating.

The energy conservation equation for the matter which is strictly residing within 
our brane is given by: $\dot \rho+ 3H(\rho+p)=0$. This has an obvious consequence 
for the scalar field dominating the early Universe during the inflationary phase. 
It has been pointed out in Refs.~\cite{binetruy,maartens} that inflation 
is well supported by $\rho^2$ contribution because of the dominance of the friction 
term leads to many e-foldings of inflation. For our purpose it is the last 
$50-60$ e-foldings of inflation should be sufficient enough to form structures 
in the Universe. The possibility of chaotic inflation with massive inflaton field 
($V(\phi) = m^2\phi^2/2$) has been discussed in Ref.~\cite{maartens}. The density perturbation 
produced by the scalar field $\phi$ during inflation has been compared to that of the 
COBE result and it has been realized that chaotic inflation can occur for field values 
below the four dimensional Planck mass $\phi_{\rm cobe} \approx 10^2 M_{\rm p}^{1/3}
\lambda^{1/6} < M_{\rm p}$, but above the five dimensional scale $M_{5}$. The mass 
of the inflaton field has also been found to be constrained $m \approx 5\times 
10^{-5}M_{5}$, which essentially translates to $ m \approx 10^{-5} M_{\rm p}^{1/3}
\lambda ^{1/6}$ from Eq.~(\ref{main1}). Hence, for $\lambda \approx 
{\cal O}({\rm GeV})^{4}$, the mass could be $m \sim {\cal O}(10){\rm GeV}$, and,
$\phi_{\rm cobe} \sim {\cal O}(10^8){\rm GeV}$. Thus the scale of inflation is 
determined by the brane tension and depending on its value inflation could 
take place at extremely low scale. One of the most important consequence of 
having inflation at a low scale is the low reheat temperature and various 
other physical implications which we will describe next.

It is known to us that inflation leads to extremely cold Universe because the entropy
generated before and during inflation redshifts away, thus it is necessary to attain 
thermalization at a scale above the nucleosynthesis scale to preserve the 
successes of the Big Bang model. We notice, after the end of inflation the scalar 
field begins oscillating coherently at the bottom of the potential, and for the 
massive inflaton the average pressure density vanishes during the oscillations,
thus leading $\rho_{\phi} \propto a^{-3}$, where $a$ is the scale factor. If we 
denote $\rho_{\phi {\rm i}}$ and $a_{\rm i}$ as the inflaton
energy density and the scale factor at the beginning of the coherent oscillations,
then the Hubble expansion is given by 
$ H^2(a)\approx ({8\pi}/{3M_{\rm p}^2})({\rho_{\phi {\rm i}}^2}/{2\lambda})({a_{\rm i}}/{a})^6$ 
If the decay rate of the inflaton is denoted by $\Gamma_{\phi}$, then equating $H(a)$ to 
$\Gamma_{\phi}$ leads to an expression for the scale 
factors. If we assume reheating occurs with the energy density in radiation 
$\rho_{\rm r}=\left(\pi^2/30\right)g_{*}T_{\rm rh}^4$, where $g_{*}$ is the 
relativistic degrees of freedom then the reheat temperature $T_{\rm rh}$ is given by
\begin{eqnarray}
\label{main4}
T_{\rm rh} \approx \left(\frac{\Gamma_{\phi}M_{\rm p} \sqrt{\lambda}}{g_*}\right)^{1/4}
\sim \left(10^{-5/4}M_{\rm p}^{1/3}\lambda^{1/6}\right)
\left(\frac{\alpha}{g_*}\right)^{1/4}\, \nonumber \\
\approx  10^{15/4}m_{\phi}\left(\frac{\alpha}{g_*}\right)^{1/4}\,,
\end{eqnarray}
where we have assumed massive boson decay $\Gamma_{\phi} \approx \alpha m_{\phi}$ with a Yukawa
coupling $\alpha$, and, while deriving the last expression in Eq.~(\ref{main4}), we 
have taken $m_{\phi} \approx 10^{-5}M_{\rm p}^{1/3}\lambda^{1/6}$. We see that the reheat
temperature is proportional to the mass of the inflaton.
For the brane tension $\lambda \sim {\cal O}(1){\rm GeV}$, reheat temperature could be
$T_{\rm rh} \approx {\cal O}(10^{3}) {\rm GeV}$, assuming $g_* \sim {\cal O}(100)$
and $\alpha \approx 0.01$. However, the reheat temperature is 
always more than the brane tension. This is a direct consequence of 
inflation occurring at low scales. Inflation at such a scale is desirable 
from the point of view of nucleosynthesis which we briefly describe here.

If we believe that supersymmetry is needed to solve the hierarchy 
between the electro-weak scale
and the four dimensional Planck mass then the gravitino mass must be no higher 
than $\sim 1$ TeV. Since we
know that gravitino coupling to matter is Planck mass suppressed, the life time of gravitino
at rest is quite long $\tau_{3/2} \sim M_{\rm p}^2/m_{3/2}^{3} 
\sim 10^{5}\left(m_{3/2}/{\rm TeV}\right)^{-3} {\rm sec}$ \cite{cline}. If the gravitino
decays to either gauge bosons and its gaugino partner, or, if it decays to energetic photons,
synthesis of light elements can be in danger by changing the number density of baryon to 
photon ratio required for a successful nucleosynthesis. However, if the Universe thermalizes 
at a temperature which is as low as ${\cal O}(10^3)$ GeV, the thermal production of 
gravitinos is also suppressed, but gravitinos could also be produced non-perturbatively
during preheating \cite{anupam}, which we do not consider here. The thermal production
of gravitino usually involves $2 \rightarrow 2$ processes involving gauge bosons and 
gauginos during reheating. In the brane-world scenario it is very likely that the bulk
is also supersymmetric and in that case there is a possibility to excite the kaluza-Klein
gravitino modes. At the Planck brane these modes are coupled to the matter field with a
Planck mass suppressed interactions and it could be very interesting to analyse them 
separately, there has been some discussion upon localization of the zero mode gravitino
in this
context \cite{oda}. In this paper we do not study exciting the kaluza-Klein gravitino
modes, however, if the formal approach to study them becomes clear then it is worth
investigating this issue separately, because they are likely to increase the 
gravitino abundance and thus likely to pose a bigger challenge to nucleosynthesis.

In order to study the gravitino abundance by strictly assuming that there is
no gravitino contribution from the bulk, we need to study the Boltzmann equation for 
the gravitino number density $n_{3/2}$ in $3+1$ dimensions \cite{kolb}.
\begin{eqnarray}
\label{imp0}
\frac{dn_{3/2}}{dt}+3Hn_{3/2}=\langle\Sigma_{\rm tot} v_{\rm rel}\rangle n^2_{\rm rad}
-\frac{m_{3/2}}{\langle E_{3/2}\rangle}\frac{n_{3/2}}{\tau_{3/2}}\,,
\end{eqnarray}
where $\langle ...\rangle$ represents thermal average, $n_{\rm rad}$ is the number 
density of relativistic particles $n_{\rm rad}\propto T^3$, $v_{\rm rel}$ is the 
relative velocity of the scattering radiation which in our 
case $\langle v_{\rm rel}\rangle =1$,
and the factor $m_{3/2}/\langle E_{3/2}\rangle$ is the average Lorenz factor. We notice
that in radiation era the non-conventional brane cosmology gives the following Hubble rate
of expansion:
\begin{eqnarray}
H \approx \left(\frac{4\pi^5}{3}\right)^{1/2}\frac{g_{*}}{30}\frac{T^4}{\sqrt{\lambda}M_{\rm p}}\,.
\end{eqnarray}
In supersymmetric version $g_{*} \sim 300$ provided the reheat temperature is more than the masses 
of the superpartners. It is worth mentioning that the scale factor during radiation era
follows $ a(t) \propto t^{1/4}$, which is contrary to the standard Big Bang scenario where 
$a(t) \propto t^{1/2}$. However, we must not forget that the derivation is based on the fact 
that we are in a regime where $\rho > 2\lambda$. In Eq.~(\ref{imp0}), after the 
end of inflation the first term in the right-hand-side dominates the second. If we assume 
adiabatic expansion of the Universe $a\propto T^{-1}$, then we can 
rewrite Eq.~(\ref{imp0}) as $Y_{3/2}=(n_{3/2}/n_{\rm rad})$. We yield
${dY_{3/2}}/{dT} \approx -({\langle \Sigma_{\rm tot}\rangle n_{\rm rad}}/{HT})$.
We notice that we can integrate the temperature dependence from this equation, and,
we mention here that the above expression is exactly the same as in the standard 
Big Bang case \cite{kolb}.
However,  this equation does not produce the correct value of $Y_{3/2}$,
since the true conserved quantity is the entropy per comoving volume. In our case if we assume 
the gravitinos do not 
decay within the time frame we are interested in, then we may be able to get the abundance 
expression at two different temperatures
\begin{eqnarray}
Y_{3/2}(T) \approx \frac{g_{*}(T)}{g_{*}(T_{\rm rh})}\frac{n_{\rm rad}(T_{\rm rh})
\langle \Sigma_{\rm tot}\rangle}{H(T_{\rm rh})}\,.
\end{eqnarray}
Here we  assume that the initial abundance of gravitinos at $T_{\rm rh}$ is known to us, and
the dilution factor $g_{*}(T)/g_{*}(T_{\rm rh})$ takes care of the decrease in the
relativistic degrees of freedom. The total cross-section $\Sigma_{\rm tot} 
\propto 1/M_{\rm p}^2$, and $n_{\rm rad}(T_{\rm rh}) \propto T_{\rm rh}^3$, we finally 
get an expression for the gravitino abundance at temperature $T$
\begin{eqnarray}
\label{life}
Y_{3/2}(T \ll 1{\rm MeV}) \approx 10^{-3} \frac{\sqrt{\lambda}}{T_{\rm rh} M_{\rm p}}\,.
\end{eqnarray}
The above expression is an important one and
now we are in a position to estimate the abundance for gravitinos. First of all we mention
that the abundance equation is in stark contrast to the conventional one $Y_{3/2} \approx 10^{-2}
(T_{\rm rh}/M_{\rm p})$, where the reheat temperature appears in the numerator rather than
in denominator. If we assume that after their creation during reheating their number density 
is conserved, then for $T_{\rm rh} \approx 10^{3} {\rm GeV}$ and 
$\lambda \approx (1{\rm GeV})^4$, we get an extremely small abundance of 
gravitinos $Y_{3/2} \approx 10^{-25}$.
However, for similar reheat temperature, the conventional Big Bang cosmology would predict
the abundance $Y_{3/2} \approx 10^{-18}$. Thus we find extremely low abundance of 
gravitinos in our case. However, there is a word of caution. The abundance depends 
on the brane tension and it can be evaluated at ease that increase in brane tension 
leads to increase in the mass of the inflaton and also the reheat temperature.
This eventually leads to extremely high abundance of gravitinos during
reheating compared to the ordinary Big Bang case. This could be a potent problem for
intermediate range five dimensional Planck mass, which is a common feature in 
M-theory compactifications. So, all is not well with the brane cosmology, however, for 
small brane tensions, $\rho^2$ contribution to the Friedmann equation could be beneficial.
In order to be a successful candidate for small brane tensions, the   
issue of baryogenesis becomes very important and this is the discussion we follow next.

An important mechanism for generating baryon asymmetry is through the decay of sfermion
condensate proposed in Ref.~\cite{ad}, known as AD mechanism.
Let us consider sfermion condensate denoted by $\psi$ and a simple potential for
$\psi$ which is lifted by breaking supersymmetry at a suitable scale 
$V \approx \tilde m ^2 \psi^2 $,
where $\tilde m$ is related to the supersymmetry breaking scale. A large baryon asymmetry can be
generated if there is a baryon number violating operator, such as $\langle A \rangle \neq 0$.
The baryon number density stored in the sfermion oscillations is given by \cite{ad}
\begin{eqnarray}
\label{ratio}
n_{\rm B}=\epsilon \left(\frac{\psi_{0}^2}{M_{\rm G}^2}\right)\frac{\rho_{\psi}}{\tilde m}\,,
\end{eqnarray}
where $\psi_0$ is the initial amplitude of the sfermion oscillations, $M_{\rm G}$ can be 
assumed to be an intermediate scale, this could be supersymmetric grand unification scale. 
$\epsilon(\psi_{0}^2/M_{\rm G}^2)$ is the net baryon number generated by the decay of $\psi$.
As we know in general that the inflaton begins oscillating when
$H \sim m_{\phi}$ at $a=a_{\phi}$ and oscillations of the sfermion begin 
quite late when $H\sim \tilde m$ at $a=a_{\psi}$. One of the most important condition
to realize the AD baryogenesis is that the thermalization due to the
decay products of the inflaton field must take place after the decay of the AD field, and, 
$\rho_{{\rm r}\phi} > \rho_{\psi}$, where $\rho_{{\rm r}\phi}$ is the energy density in radiation
after the inflaton decay. This is mainly required to prevent washing out the baryon asymmetry.
This tells us that the inflaton should decay very slowly and possibly via gravitational interactions,
however, if this is so, then most probably the Universe would undergo transition from 
non-conventional to the standard one while the process of reheating. This will happen when
$\rho_{\phi}\approx m_{\phi}^2\phi^2(a_{\phi}/a)^3 \sim \lambda$ at
$a =a_{\lambda} =({m_{\phi}^2\phi^2}/{\lambda})^{1/3}a_{\phi}$
We picturize a situation where the Universe began with a 
non-conventional cosmology, then after the end of inflation the inflaton begins oscillating, 
but the Universe is still non-conventional. When the Hubble parameter drops to a 
value $H \sim \tilde m$ the oscillations in the AD field begins and at this time also the 
Universe is non-conventional.
However, soon after oscillations in the AD field is induced, the transition from  
non-conventional to the standard cosmology paves its way. Since the mass of the AD field 
is $\tilde m \sim m_{3/2} < m_{\phi}$
small compared to the mass of the inflaton, the oscillations in the AD field begin after 
the inflaton oscillations. This can be estimated by taking $H\sim \tilde m$. Since this happens
when the Universe is non-conventional; $H \approx (m_{\phi}^2\phi^2/M_{\rm p}\sqrt{\lambda})
(a_{\phi}/a)^3 \sim \tilde m$. We can estimate the scale factor when this happens
$a=a_{\psi} =({\sqrt{\lambda}}/{M_{\rm p}\tilde m})^{1/3}a_{\lambda}$.
It can be verified easily that $a_{\psi}<a_{\lambda}$.
However, this restricts the five dimensional Planck mass $M_{5}< 10^{14}$ GeV.
After $a_{\lambda}$ the cosmology becomes the standard one and the Hubble 
rate is given by $H\propto \sqrt{\rho}/M_{\rm p}$. In our set-up the inflaton decays when the
Universe is already in the standard cosmology, thus we can estimate the scale factor 
when this happens by equating the Hubble parameter to the decay rate of the inflaton; 
$ H \approx (m_{\phi}\phi/M_{\rm p})(a_{\phi}/a_{\lambda})^{3/2} 
(a_{\lambda}/a)^{3/2} \sim \Gamma_{\phi}=(m_{\phi}^3/M_{\rm p}^2)$. Notice that the decay rate 
of the inflaton is via the gravitational coupling. This yields
$a=a_{{\rm d}{\phi}}=({\lambda M_{\rm p}^2}/{m_{\phi}^6})^{1/3}a_{\lambda}$.
It can be verified that $a_{\phi}< a_{\psi} < a_{\lambda}< a_{{\rm d}\phi}$, this 
also requires to use the constraint on the mass of the inflaton; $m_{\phi} \sim
10^{-5}M_{\rm p}^{1/3}\lambda^{1/6}$ \cite{maartens}. 
During the oscillations of the AD field, the energy density decreases in the same fashion as in the  
case of inflaton. We can estimate the energy density in the AD field by
$\rho_{\psi}=\tilde m^2\psi_{0}^2({a_{\psi}}/{a})^3 =({\tilde m\sqrt{\lambda}\psi_{0}^2}
/{M_{\rm p}})({a_{\lambda}}/{a})^3 $. It can be easily verified that 
for larger $\psi_{0}$, $\Gamma_{\phi}/\Gamma_{\psi} >1$ for the sfermion decay rate $\Gamma_{\psi}
\sim (\tilde m^3/\psi^2)$ \cite{ellis2}. However, in this case an important factor is that  
thermalization due to the decay of the inflaton field must happen after the full decay of the AD
field.     

Once the Universe becomes radiation dominated, the energy density of the relativistic decay products
of the inflaton can be given by
$\rho_{{\rm r}\phi}=({m_{\phi}^6}/{M_{\rm p}^2})({a_{{\rm d}\phi}}/{a_{\lambda}})^4
({a_{\lambda}}/{a})^{4}=({\lambda^{4/3}M_{\rm p}^{2/3}}/{m_{\phi}^2})
({a_{\lambda}}/{a})^4$, and, the Hubble parameter is given by
$H=({\lambda^{2/3}}/{m_{\phi}M_{\rm p}^{2/3}})({a_{\lambda}}/{a})^{2}$.
Now we must estimate when the AD field decays, following Refs.~\cite{ellis1} and \cite{ellis2} 
we equate $H\sim \Gamma_{\psi} \equiv \tilde m^3/\psi^2$.
This takes place when the scale factor is given by
$a = a_{{\rm d}\psi}=({\lambda^{7/6}\psi_{0}^2}/{\tilde m^4 m_{\phi} M_{\rm p}^{5/3}})^{1/5}
a_{\lambda}$.
It can be verified that $\rho_{{\rm r}\phi}(a_{{\rm d}\psi}) > \rho_{\psi}(a_{{\rm d}\psi})$.
Now we have to make sure that the thermalization of the inflaton field happens after the decay of the
the AD field. For that we need to estimate the thermalization rate of the inflaton field.
Following the arguments given in Refs.\cite{ellis1} and \cite{ellis2} we get
$\Gamma_{\rm T}\sim n_{\phi}\sigma \sim m_{\phi}\phi^2({a_{\phi}}/{a})^3 \sim
({\alpha^2}/{m_{\phi}^2})({a}/{a_{{\rm d}\phi}})^2
\sim \alpha^2({\lambda^{1/3}m_{\phi}}/{M_{\rm p}^{4/3}})({a_{\lambda}}/{a})$,
where $n_{\phi}$ is the number density of the relativistic particles, $\sigma$ is the cross-section
and $\alpha$ is the fine structure constant. The thermalization of the Universe occurs when
$\Gamma_{\rm T} \sim H$, and,
$a_{\rm T}=\alpha^{-2}({\lambda^{1/3}M_{\rm p}^{2/3}}/{m_{\phi}^2})a_{\lambda}$.
At this point we can also check that $a_{{\rm d}\psi} < a_{\rm T}$ for $m_{\phi} \sim 10^{-5}M_{5}$, 
and $\alpha \sim 10^{-3/2}$. The condition is satisfied for any reasonable 
value of $\psi_0$ less than the four dimensional Planck mass.

At $a_{\rm T}$ we can compute the final baryon to entropy ratio given by \cite{ad}. We also
have to compute the entropy, which is given by: $s=(\rho_{{\rm r}\phi}(a_{\rm T}))^{3/4} 
\approx ({\alpha^{6}m_{\phi}^{9/2}}/{M_{\rm p}^{3/2}})$
and finally the baryon to entropy ratio can be given by 
\begin{eqnarray}
\label{final}
\frac{n_{\rm B}}{s}=\frac{\epsilon \psi_{0}^4 m_{\phi}^{3/2}}{M_{\rm G}^2\sqrt{\lambda}
M_{\rm p}^{3/2}}\equiv \frac{\epsilon \psi_{0}^4 m_{\phi}^{3/2}}{M_{\rm G}^2 
M_{5}^3 M_{\rm p}^{1/2}}\,.
\end{eqnarray}
It is noticeable that the baryon to entropy ratio does not depend on $\tilde m$. However, it does 
depend on the brane tension and the initial amplitude of the AD field oscillations. The last step
in the above equation has been been expressed in terms of the five dimensional Planck mass.
For an example, we may take $M_{\rm G} \sim 10^{15}$ GeV, $m_{\phi}\sim 10^{-5}M_{5}$, we get
an estimation of the initial amplitude of oscillations in the AD field
$\psi_{0}=({10^{37}}/{\epsilon})^{1/4}({M_{5}}/{\rm GeV})^{3/8}~{\rm GeV}$,
where we have taken the observed baryon to entropy ratio to be $n_{\rm B}/s \sim 10^{-10}$.
It is evident that the value of $\psi_{0}$ is more than $\phi_{\rm COBE} \approx 10^{2}M_{5}$.
However, for smaller values of $M_{5}$ the amplitude could be comparable to $\phi_{\rm COBE}$.
In that case, situation could be different. Here we have implicitly assumed that the AD field
decays after the decay of the inflaton. For smaller values of $\psi_{0}$, the situation could be
reversed, in that case the AD field would decay before the inflaton decay. In such a case,
the entropy produced would be simply given by the inflaton decay and we do not have to bother about 
actual thermalization of the relativistic particles.

Here we summarise by saying that the brane-world cosmology differs quite a lot in their predictions
from the standard cosmology. Here we have looked upon two issues, the gravitino abundance and
the baryogenesis. Other interesting issues should also be taken into account and work in this 
direction is in progress.


ACKNOWLEDGEMENTS: The author is supported by INLAKS foundation. The author is thankful to 
Andrew Liddle for fruitful discussion.


\vspace*{-0.7truecm}
\begin{references}
\vspace*{-1.6truecm}

\bibitem{horava} P. Horava and E. Witten, Nucl. Phys. B {\bf 460} (1996) 506; Nucl. Phys. B
   {\bf 475} (1996) 94.

\bibitem{lukas0}A. Lukas, B. A. Ovrut, K. Stelle and D. Waldram, Phys. Rev. D {\bf 60}
 (1999) 086001; J. E. Ellis, Z. Lalak, S. Pokorski, W. Pokorski, Nucl. Phys. B {\bf 540}
 (1999) 149.

\bibitem{lukas1}A. Lukas, B. A. Ovrut, D. Waldram, Phys. Rev. D {\bf 61} (2000) 023506.



\bibitem{binetruy}P. Binetruy, C. Daffayet and D. Langlois, Nucl. Phys. B {\bf 565}
 (2000) 269; P. Binetruy, C. daffayet, U. Ellwanger and D. langlois, Phys. Lett. B {\bf 477}
 (2000) 269.


\bibitem{ad}I. Affleck and M. Dine, Nucl. Phys. B {\bf 249} (1985) 361.

\bibitem{lisa} L. Randall and R. Sundrum, Phys. Rev. Lett. {\bf 83} (1999) 3370.



\bibitem{bagger}R. Altendorfer, J. Bagger and D. Nemeschansky, hep-th/0003117;
   A. Falkowski, Z. Lalak and S. Pokorski, Phys. Lett. B {\bf 491} (2000) 172.

\bibitem{kallosh}R.  Kallosh and A. Linde JHEP, 0002, (2000) 005;
E. Bergshoeff, R. Kallosh and A. Van Proeyen, hep-th/0007044.

\bibitem{stelle}M.J. Duff, J. T. Liu and  K. S. Stelle, hep-th/0007120. 

\bibitem{duff} M.J. Duff, J. T. Liu and  W.A. Sabra, hep-th/0009212. 

\bibitem{cgs}J. M. Cline, C. Grojean and G. Servant, Phys. Rev. Lett. {\bf 83} (1999) 4245.

\bibitem{lisa1} L. Randall and R. Sundrum, Phys. Rev. Lett. {\bf 83} (1999) 4690.

\bibitem{maartens}R. Maartens, D. Wands, B. A. Bassett and I. P. C. Heard, Phys. Rev. D {\bf 62}
   (2000) 041301; E. J. Copeland, A. R. Liddle and J. E. Lidsey, astro-ph/0006421.

\bibitem{kolb} E. W. Kolb and M. Turner, {\it The Early Universe}, Addison-Wesley, 1993; 
M. Kawasaki, and T. Moroi, Prog. Theor. Phys. {\bf 93}, (1995) 879, T. Moroi, P.hD. Thesis, 
hep-ph/9503210. 

\bibitem{cline} J. Cline and S. Raby, Phys. Rev. D {\bf 43} (1991) 1781.

\bibitem{anupam} A. L. Maroto and A. Mazumdar, Phys. Rev. Lett. {\bf 84} (2000) 1655;
 R. Kallosh, L. Kofman, A. Linde and A. V. Proyen, Phys. Rev. D {\bf 61} (2000) 103503;
 G. F. Giudice, A. Riotto and I. Tkachev, JHEP 9908 (1999) 009.

\bibitem{oda} I. Oda, hep-th/0008134.

\bibitem{ellis1}J. Ellis, D. V. Nanopoulos and K. A. Olive, Phys. Lett. B {\bf 184}, (1987) 37.

\bibitem{ellis2}J. Ellis, K. Enqvist, D. V. Nanopoulos and K. A. Olive, Phys. Lett. B
{\bf 191}, (1987) 343.
\end{references}
\end{document}